\documentclass{iau} 
\usepackage{graphicx}
\title[$N$-body chaos and phase-space transport in simulations]{$N$-body chaos, phase-space transport and relaxation in numerical simulations}
\author[P.\ Di Cintio \& L.\ Casetti]{Pierfrancesco Di Cintio$^{1,2}$ \and Lapo Casetti$^{2,3,4}$}
\affiliation{$^1$IFAC-CNR, Via Madonna del piano 10,
I-50019, Sesto Fiorentino (FI), Italy \\ email: {\tt p.dicintio@ifac.cnr.it} \\[\affilskip]
$^2$INFN, Sezione di Firenze, via G.\ Sansone 1,
I-50019, Sesto Fiorentino (FI), Italy\\
$^3$Dipartimento di Fisica e Astronomia, Universit\`a di Firenze,\\ via G.\ Sansone 1,
I-50019, Sesto Fiorentino (FI), Italy\\
$^4$INAF-Osservatorio astrofisico di Arcetri, largo E.\ Fermi 5, I-50125, Firenze, Italy}
\pubyear{2019}
\volume{351}  
\jname{Star Clusters: From the Milky Way to the Early Universe}
\editors{A. Bragaglia, M.B. Davies, A. Sills \& E. Vesperini, eds.}
\begin{document}
\maketitle
\begin{abstract}
Using direct $N$-body simulations of self-gravitating systems we study the dependence of dynamical chaos on the system size $N$. We find that the $N$-body chaos quantified in terms of the largest Lyapunov exponent $\Lambda_{\rm max}$ decreases with $N$. The values of its inverse (the so-called Lyapunov time $t_\lambda$) are found to be smaller than the two-body collisional relaxation time but larger than the typical violent relaxation time, thus suggesting the existence of another collective time scale connected to many-body chaos. 
\keywords{stellar dynamics, galaxies: kinematics and dynamics, methods: n-body simulations, diffusion.}
\end{abstract}
\firstsection 
\section{Introduction}
The dynamics of $N$-body gravitational systems, due to the long-range nature of the $1/r^2$ force, is dominated by {\it mean-field} effects rather than by inter-particle collisions for large $N$. Due to the extremely large number of particles it is often natural to adopt a description in the continuum ($N\to\infty$, $m\to 0$) collisionless limit in terms 
of the Collisionless Boltzmann Equation (CBE, see e.g.\ \cite{bt08}) for the single-particle phase-space distribution function $f(\mathbf{r},\mathbf{v},t)$
\begin{equation}\label{vlasov}
\partial_t f+\mathbf{v}\cdot\nabla_{\mathbf{r}}f+\nabla\Phi\cdot\nabla_{\mathbf{v}}f=0,
\end{equation}
coupled to the Poisson equation $\Delta\Phi(\mathbf{r})=4\pi G\rho(\mathbf{r})$. In many real self-gravitating systems, such as elliptical galaxies, $N$ is large ($N\approx 10^{11}$), and the two-body relaxation time $t_{2b}\propto Nt_*/\log N$ associated to collisional processes \cite{1943ApJ....97..255C} is much larger than the age of the universe. The typical states of such systems are therefore idealized as collisionless equilibria of Eq.\ (\ref{vlasov}), approached via the mechanism of violent relaxation first suggested by \cite{lyn69}, on a time scale of the order of the average crossing time $t_*$.\\
\indent The question whether the continuum limit is effectively meaningful, and how the intrinsic discrete nature (see e.g.\ \cite{1980PhR....63....1K}) of collisionless systems affects their relaxation to (meta-)equilibrium is in principle still open.\\
\indent Using arguments of differential geometry \cite{gurz86}, conjectured the existence of another relaxation time $t_*<\tau<t_{2b}$, linked to the inverse of the  exponentiation rate of phase-space volumes, and scaling as $N^{1/3}$ with the system size (see also \cite{vesp92}).
\cite{kandrup01,kandrup02,kandrup03,kandrup04} studied the ensemble properties of tracer orbits in frozen $N$-body potentials (i.e., the potential generated by $N$ {\it fixed} particles distributed according to different densities $\rho(r)$). They found that even at large $N$ the discreteness effects are not negligible in models with both integrable and non-integral continuum limit counterparts. Moreover they find that the exponentiation rate of the phase-space volume of initially localized clumps of particles is compatible with the mean Lyapunov exponent associated to the particles. The latter was found to be a slightly increasing function of $N$. In parallel \cite{hem} studied the $N$-body chaos in self-consistent Plummer models as a function of $N$, quantified by means of the exponentiation rate of a subset of the $6N$ phase-space volume considering only the position part and finding again a slightly increasing degree of chaoticity.\\
\indent More recently, \cite{beraldo19}, using information entropy arguments, suggested the existence of another relaxation scale, associated to discreteness effects in $N$-body models, that scales as $N^{-1/6}$.\\ 
\indent Here we revisit this matter further by studying the dynamics of individual tracer particle in frozen and self-consistent equilibrium models as well as the Lyapunov exponents of the full $N$-body problem. 
\section{Methods}
In both frozen and active models we consider the spherically symmetric Plummer density profile
\begin{equation}\label{plummer}
\rho(r)=\frac{3}{4\pi}\frac{Mr_c^2}{(r_c^2+r^2)^{5/2}},
\end{equation}
with total mass $M$ and core radius $r_c$. In order to generate the velocities for the active simulations, we use the standard rejection technique to sample the anisotropic equilibrium phase-space distribution function $f$ obtained from $\rho$ with the standard \cite{edd16} integral inversion.
Throughout this work we assume units such that $G=M=r_c=1$, so that the dynamical time $t_*=\sqrt{r_c^3/GM}$ and the scale velocity $v_*=r_c/t_*$ are also equal to unity. Individual particle masses are then $m=1/N$. For the numerical simulations we use a standard fourth-order symplectic integrator with fixed time-step $\Delta t=5\times 10^{-3}$ to solve the particles' equations of motion and the associated tangent dynamics used to evaluate the maximal Lyapunov exponent with the standard \cite{bgs} method, as the large $k$ limit of
\begin{equation}
\Lambda_{\rm max}(t)=\frac{1}{k\Delta t}\sum_k\ln\left|\frac{W(k\Delta t)}{d_0}\right|.
\end{equation}
In the above equation $W$ is the norm of the $6N$- or 6-dimensional tangent vector for a self-consistent $N$-body simulation and an individual particle orbit, respectively. In both cases $d_0=W(t=0)$.  
\begin{figure}[ht!]
\begin{center}
 \includegraphics[width=3in]{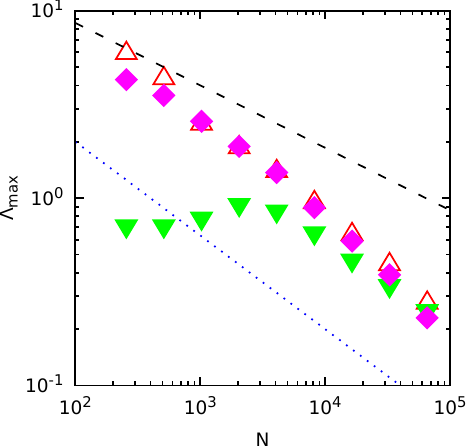} 
 \caption{Maximal Lyapunov $\Lambda_{\rm max}$ exponent as a function of $N$ for the full $N$-body problem (upward triangles) and for a single tracer orbit in the time dependent potential of all the others (downward triangles). Average maximal Lyapunov exponent for an ensemble of non-interacting particles sampled from an isotropic Plummer model propagated in a frozen Plummer model (diamonds).}
   \label{fig1}
\end{center}
\end{figure}
\section{Results and discussion}
\begin{figure}[ht!]
\begin{center}
 \includegraphics[width=2.9in]{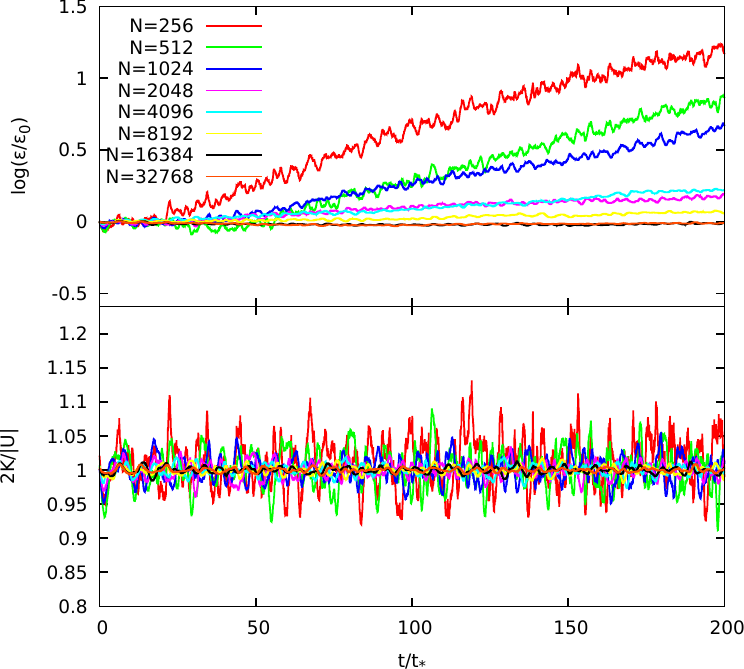} 
  \includegraphics[width=2.2in]{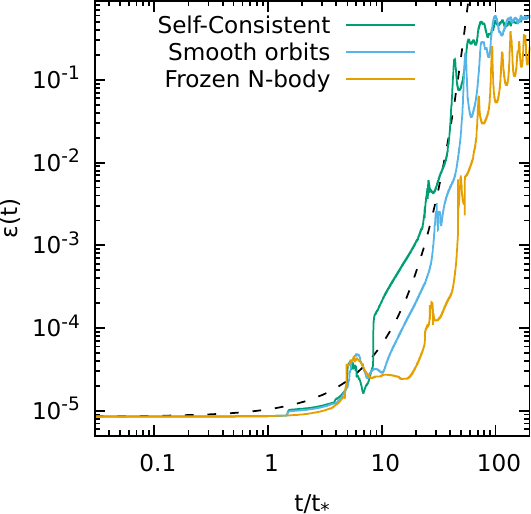} 
 \caption{Top left: evolution of the normalized collective emittance for an  isotropic Plummer model with different values of $N$. Bottom left: evolution of the virial ratio $2K/|U|$. Right: evolution of the emittance of an initially localized cluster in the $N=16384$ case.}
   \label{fig2}
\end{center}
\end{figure}

\cite{dcs} have computed the largest Lyapunov exponents $\Lambda_{\rm max}$ for the full $N$-body problem and for several tracer orbits for different system sizes $N$,  finding that in the first case $\Lambda_{\rm max}$ is a decreasing function of $N$ with a slope between $-1/3$ and $-1/2$ (upward triangles in Fig. \ref{fig1}), contrary to previous numerical estimates (see e.g.\ \cite{hem}). In the second case, when following a single particle in the potential of the others, $\Lambda_{\rm max}$ is almost constant in both active and frozen potentials for tightly bound particles, while decreases for less bound particles with different power-law trends as function of the binding energy.\\ 
\indent Curiously, when measuring the average largest Lyapunov exponent of a system of non-interacting particles randomly sampled from a self-consistent isotropic system and propagated in a frozen realization of the latter, $\langle\Lambda_{\rm max}\rangle$ has the same scaling with $N$ (and remarkably close values), as shown by the diamonds in Fig. \ref{fig1}.\\
\indent Here we compute the so-called emittance, a quantity defined in the context of charged particles beams (see e.g.\ \cite{kandrup04} and references therein) associated to the ``diffusion'' in phase-space volume, defined by
\begin{equation}
\epsilon=(\epsilon_x\epsilon_y\epsilon_z)^{1/3},\quad \epsilon_i=\sqrt{\langle r_i^2\rangle\langle v_i^2\rangle-\langle r_iv_i\rangle^2},
\end{equation}
where $\langle...\rangle$ denotes ensemble averages. We have evaluated $\epsilon$ for the full $N$-body problem, finding that the global system's emittance is almost conserved for large $N$, while it appears to grow with time (although the model remains in virial equilibrium) for $N\lesssim 2000$, as shown in Fig.\ \ref{fig2} (left pannels).
When we consider instead the emittance of an initially localized cluster of tracers (see Fig. \ref{fig2}, right pannel) in an active $N-$body system, we observe that the latter has (independently of $N$) an increasing beahaviour with time, roughly proportional to $\epsilon_0\exp(\langle\Lambda_{\rm max}\rangle t)$ (dashed line), where $\epsilon_0$ is its value at $t=0$ and $\langle\Lambda_{\rm max}\rangle$ is the mean maximal Lyapunov exponent of the cluster. We have repeated the same numerical experiment for frozen $N$=body potentials and for time dependent potentials generated by non-interacting particle in a smooth Plummer model finding the same behaviour (See again Fig.\ \ref{fig2}). Moreover, in direct $N$-body simulations, when we compare the evolution of $\epsilon$ for a subset of active particles with that of the same cluster of non-interacting particles, we find that for large $N$ $\epsilon$ grows in the same fashion for both tracers and active particles. In turn, $\epsilon$ retains generally lower values for initially localized clusters in the frozen case, as a consequence of the conservation of energy and smaller fluctuations of the angular momentum in the latter case (see \cite{dcs}).\\
\indent All these results lead us to conclude that the discreteness effects and the intrinsic chaoticity of the $N$-body problem may be linked to another effective relaxation scale shorter than $t_{2b}$. Moreover, it also appears that the conclusions arising from the study of frozen $N$-body models can not be easily extended to real self-consistent systems. The results on the effect of the discreteness chaos on collective instabilities such as the radial-orbit instability (ROI) will be published elsewere [\cite{dcs2}].

\end{document}